\documentclass{osa-article}

\journal{oe}

\articletype{Research Article}

\usepackage{lineno}

\begin{document}
	
	\title{Physics-based basis functions for low-dimensional representation of the refractive index in the high energy limit}
	
	\author{Saransh Singh,\authormark{1,*} and K. Aditya Mohan\authormark{1}}
	
	\address{\authormark{1}Lawrence Livermore National Laboratory, Livermore, CA 94568, USA\\}
	
	\email{\authormark{*}saransh1@llnl.gov} 
	
	
	
	\begin{abstract}
		The relationship between the refractive index decrement, $\delta$, and the real part of the atomic form factor, $f^\prime$, is used to derive a simple polynomial functional form for $\delta(E)$ far from the K-edge of the element. The functional form, motivated by the underlying physics, follows an infinite power sum, with most of the energy dependence captured by a single term, $1/E^2$. The derived functional form shows excellent agreement with theoretical and experimentally recorded values. This work is useful to reduce the dimensionality of the refractive index across the energy range of x-ray radiation for efficient forward modeling and formulation of a well-posed inverse problem in propagation based polychromatic phase-contrast computed tomography.
	\end{abstract}
	
	\section{Introduction}
	Phase contrast tomography (PCT) is a non-destructive 3D characterization technique that can be used to obtain the 3D distribution of the refractive index, i.e., the real part of the sample's complex refractive index, which encodes the phase shift during x-ray interaction with the sample. This differs from the conventional absorption-based computed tomography method that reconstructs the 3D distribution of the imaginary part of the refractive index, which encodes the absorption of x-ray by the sample. PCT has benefits over conventional absorption-based computed tomography (CT) techniques for producing good contrast in samples containing low atomic number materials. However, quantitative PCT has been limited to highly monochromatic synchrotron sources. 
	
	In order to perform 3D quantitative refractive index reconstruction using poly-chromatic x-ray sources, a low-dimensional accurate representation of the energy variation of the refractive index is needed. Mathematically, the variation in the refractive index is conveniently represented using basis functions. These functions provide a low dimensional representation of the refractive index's energy dependence as only a set of coefficients to the these basis functions. This approach has previously been used in dual-energy x-ray computed tomography (DECT) to represent the linear-attenuation coefficient (LAC)~\cite{Martz2016,Azevedo2016,champley2019}. The DECT data acquired on several systems and varied energy spectra were subsequently used to reconstruct the average electron density, $\rho_e$ and average atomic number, $Z_e$, so-called SIRZ \cite{Martz2016,Azevedo2016}, SIRZ-2 \cite{champley2019} and SIRZ-3 \cite{Mohan2022} methods. It is interesting to note that the authors in Ref.~\cite{champley2019} saw a massive improvement in the relative errors for $\rho_e$ by using physically motivated basis functions for the LAC.
	This contribution lays the groundwork for quantitative PCT at the poly-energetic white beam synchrotron and lab-based x-ray systems. Using the Kramer-Kronig relationship for an analytical function, simple polynomial basis functions for the refractive index are derived. During the inverse phase reconstruction step, the reciprocal function series significantly reduce the degrees of freedom in the inverse problem, and aids in the formulation of a well-posed inverse problem. Typically, only one to two coefficients need to be determined at each pixel, making the phase reconstruction using poly-chromatic x-rays tractable.
	The paper is organized as follows: Section~\ref{sec:theory} outlines the derivation of the basis functions based on the real and imaginary part of the anomalous scattering factors \cite{Henke1993,Cromer1970,Chantler1995}. Section~\ref{sec:results} presents the interpolation results using these basis functions on tabulated values using theoretical calculations as well as experimentally recorded values. Some of the details of the calculations are detailed in Appendix sections \ref{sec:app1}, \ref{sec:Analytical}. Finally, section~\ref{sec:conclusions} concludes the paper with some final remarks and avenues for future work.
	
	\section{Theory\label{sec:theory}}
\begin{figure}[!ht]
		\centering
		\includegraphics[width=0.85\textwidth]{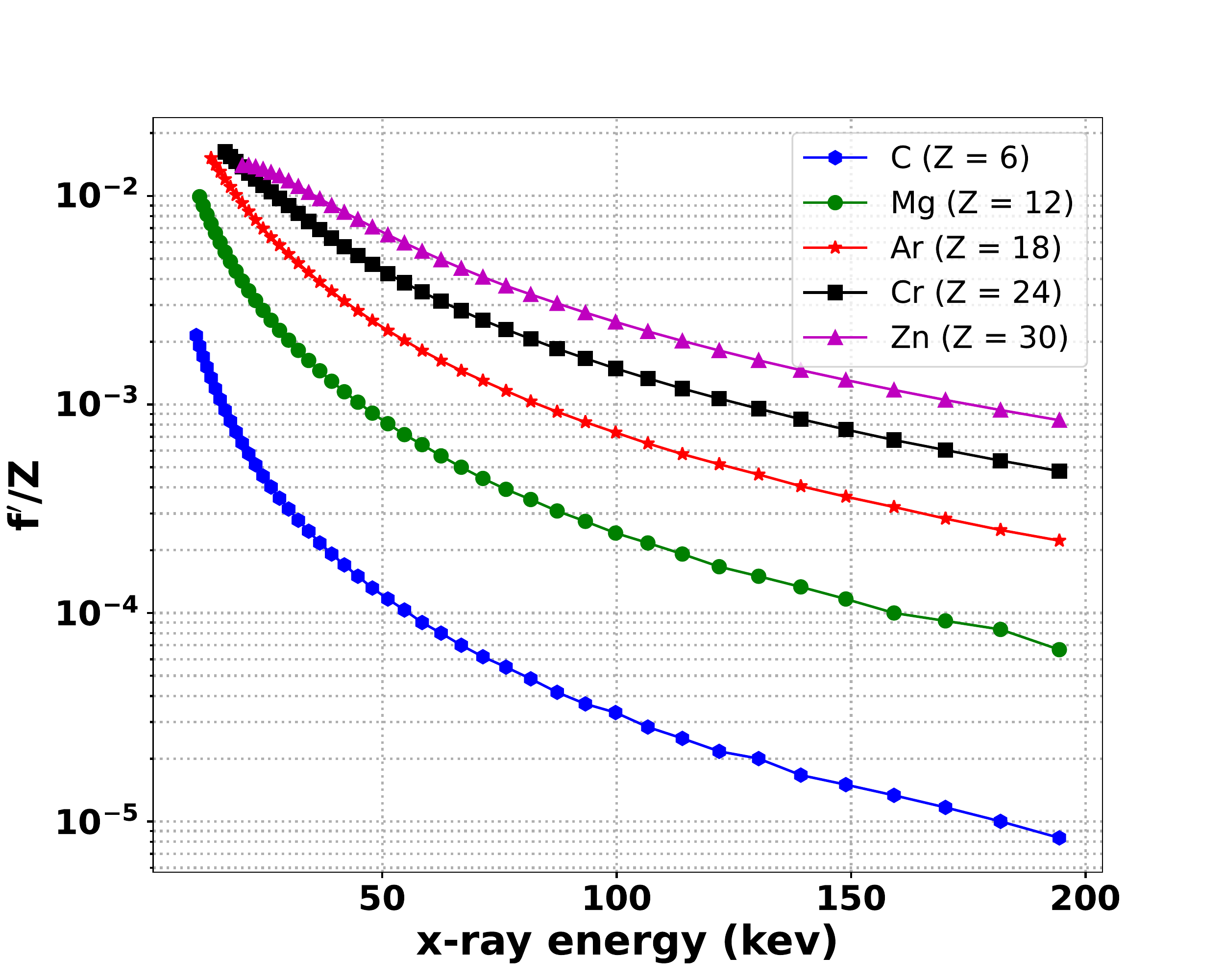}
		\caption{Semi-log plot of the ratio of real part of the anomalous scattering factor, $f^{\prime}$ with the atomic number, $Z$ as a function of the x-ray energy for a range of atomic numbers. The $f^{\prime}/Z$ ratio decreases at higher energies, implying that the accuracy of the 1/$(\hbar\omega)^{2}$ scaling for the refractive index, $\delta$ improves as the energy increases.}
		\label{fig:f_z}
	\end{figure}

	The complex refractive index of a material is a complex number, where the real part quantifies the phase shift due to propagation. In contrast, the imaginary part quantifies the attenuation in the medium. Since the refractive index only deviates from unity by a small amount, the following equation for the refractive index is typically used
	\begin{equation}
		n(\hbar\omega) = 1 - \delta(\hbar\omega) + \textrm{i}\beta(\hbar\omega).
	\end{equation}
	Here, $n$ represents the complex refractive index, $\delta$ is the deviation of the refractive index from unity and $\beta$ is the absorption coefficient which quantifies the the x-ray absorption in the medium due to the photoelectric cross-section and $\hbar\omega$ is the x-ray energy. The amount of phase shift due to x-ray of energy $\hbar\omega$ propagating in the medium is given by 
	\begin{equation}
		\phi(\hbar\omega) = \int_{s}\delta(\hbar\omega, s)\textrm{d}s.
	\end{equation}
	Here, $s$ is the x-ray path in the medium. Phase contrast tomography is the reconstruction of the refractive index decrement, $\delta$, as a function of its spatial location. In the x-ray energies used for phase contrast imaging, $\delta$ and $\beta$ are small positive numbers for most materials. For the remainder of the paper, the explicit energy dependence in for the refractive index, $\delta \equiv \delta(\hbar\omega)$, and absorption coefficient, $\beta \equiv \beta(\hbar\omega)$, will be dropped for clarity. The decrement in refractive index, $\delta$, and the imaginary part of the refractive index, $\beta$ are related to the complex anomalous atomic scattering factor, $f = f^{\prime} + \textrm{i}f^{\prime\prime}$ by the following relationship \cite{Henke1993}
	\begin{align}
		\label{eq:deltabeta}
		\delta &= \frac{2\pi \rho_{n} r_e}{(\hbar\omega)^2}\sum_{i}n_{i}(Z_{i} + f_{i}^\prime),\\ \nonumber
		\beta &= \frac{2\pi \rho_{n} r_e}{(\hbar\omega)^2}\sum_{i} n_{i}f_{i}^{\prime\prime}.
	\end{align}
	Here, $r_{e}$ is the classical electron radius, $\hbar\omega$ is the energy of the x-ray photon, $\rho_{n}$ is the number density (number of atoms per unit volume), $Z_{i}$ is the atomic number of the $i^{\textrm{th}}$ element, and $n_{i}$ is the number of atom type $i$ in the formula unit. The sum runs over different atom types in the material. At the energies of interest to phase contrast imaging ($10-200$ keV), the real part of the anomalous scattering factor for an element, $f^{\prime}$, is two to three orders of magnitude smaller than the atomic number of that element, $Z$. This is shown in Fig.~\ref{fig:f_z}. Therefore, eq.~\ref{eq:deltabeta} implies that the dominant scaling term with energy for $\delta(\hbar\omega)$ is given by $(\hbar\omega)^{-2}$.
	The real and imaginary part of the atomic form factor are related to each other by the Kramers-Kronig relationship\cite{Kronig1926}. Mathematically, this relationship guarantees that the complex scattering factor is analytic, which is required to maintain strict causality \cite{Toll1956}. The relationship is given by
	\begin{equation}
		f^{\prime} = \frac{2}{\pi}\mathcal{P}\int_{0}^{\infty}\frac{\omega^{\prime}f^{\prime\prime}(\omega^{\prime})}{\omega^{\prime 2} - \omega^{2}}\mathrm{d}\omega.
		\label{eq:kk}
	\end{equation}
	In the above equation, $\mathcal{P}(\cdot)$ represents the Principal value of the integral and $\omega, \omega^{\prime}$ represent the frequency, or conversely the energy of the x-ray. The imaginary part of the atomic form factor, $f^{\prime\prime}$ is directly related to the photoelectric cross-section, $\sigma_{\textrm{\tiny PE}}(\omega)$. This cross-section describes the interaction of the incident x-rays with the electrons in the atoms, leading to ejection of the electron from the atom. Only the photoelectric cross section is necessary and sufficient. The decrement in real part of the refractive index, $\delta$, is related to the imaginary part of the complex refractive index, $\beta$, via the Kramers-Kronig relationship. Since $\beta$ depends only on the photo-absorption cross-section, the other cross-sections for scattering and absorption such as Rayleigh scattering, Compton scattering, pair production, etc. should not be included. Including the other terms results in the linear attenuation coefficient, which is different from $\beta$.
	Following eq. 27 in Ref.~\cite{Cromer1970}, the contribution of the K-edge to the anomalous scattering factor is given by
	\begin{align}
		f^{\prime}(\hbar\omega) = \frac{1}{2\pi^{2}\alpha} &\left[ \int_{0}^{\infty}\frac{\sigma(\hbar\omega^{\prime} - \epsilon_{\text{\tiny K}})(\hbar\omega^{\prime} - \epsilon_{\text{\tiny K}})^{2} - \sigma(\hbar\omega)(\hbar\omega)^{2}}{(\hbar\omega)^2 - (\hbar\omega^{\prime} - \epsilon_{\text{\tiny K}})^{2}}\mathrm{d}\omega^{\prime}\right. \\
		&+ \left.\mathcal{P} \int_{0}^{\infty}\frac{\sigma(\hbar\omega)(\hbar\omega)^{2}}{(\hbar\omega)^2 - (\hbar\omega^{\prime} - \epsilon_{\text{\tiny K}})^{2}}\mathrm{d}\omega^{\prime} \right].
		\label{eq:fp_pe}
	\end{align}
	In the equation above, $\alpha$ is the fine structure constant, $\hbar\omega$ is the photon energy, $\epsilon_{\text{\tiny K}}$ is the K-edge of the atom and $\sigma(\hbar\omega)$ is the photoelectric cross section of the atom at photon energy of $\hbar\omega$. The contribution due to the other bound states of the electron, i.e., the L, M edges are very small at higher energies and can be safely ignored. The limits in the above integral can be made finite by performing the following substitution
	\begin{equation}
		x = \frac{-\epsilon_{\text{\tiny K}}}{\hbar\omega - \epsilon_{\text{\tiny K}}}.
	\end{equation}
	
	\begin{align}
		f^{\prime}(\hbar\omega) = \frac{1}{2\pi^2\alpha} &\left[\int_{0}^{1} \frac{\sigma(-\epsilon_{\text{\tiny K}}/x)\epsilon_{\text{\tiny K}}^2 - \sigma(\hbar\omega)(\hbar\omega)^{2}x^{2}}{x^{2}\left[x^{2}(\hbar\omega)^{2} - \epsilon_{\text{\tiny K}}^{2}\right]}~\mathrm{d}x \right. \\
		&- \left. \frac{(\hbar\omega)\sigma(\hbar\omega)}{2} \ln{\left(\frac{\hbar\omega - \epsilon_{\text{\tiny K}}}{\hbar\omega + \epsilon_{\text{\tiny K}}}\right) } \right].
		\label{eq:f+}
	\end{align}
	For light elements, the energy of incident x-rays used in phase contrast imaging is much higher than the binding energy of electrons. Therefore, $\epsilon/\hbar\omega \ll 1$ is a valid approximation. Under this approximation, the second term in eq.~\ref{eq:f+} can be approximated as
	\begin{equation}
		\frac{1}{4\pi^{2}\alpha}(\hbar\omega)\sigma(\hbar\omega) \ln{\left(\frac{\hbar\omega - \epsilon_{\text{\tiny K}}}{\hbar\omega + \epsilon_{\text{\tiny K}}} \right)} \approx  \frac{1}{4\pi^{2}\alpha}(\hbar\omega)\sigma(\hbar\omega) \left(\frac{-2\epsilon_{\text{\tiny K}}}{\hbar\omega}\right) \approx \frac{-1}{2\pi^{2}\alpha}\epsilon_{\text{\tiny K}} \sigma(\hbar\omega)
	\end{equation}
	If the functional form for the photoelectric cross-section is available, then the first term can be integrated analytically. The variation of the photoelectric cross-section is well represented by the functional form \cite{james1965,cullen1997}
	\begin{equation*}
		\sigma(\hbar\omega) = \frac{K}{(\hbar\omega)^{m}}.
	\end{equation*}
	The value of $m$ lies between $2$ and $3$ for most elements (see Fig.~\ref{fig:supp1}). Analytical integral for specific values of $m=2,2.5,3$ have been calculated in Ref.~\cite{james1965} and provided in Appendix eq.~\ref{eq:analytic_fpp}. However, $m$ in the equation above is a function of the atomic number. Therefore, the accuracy of the basis functions can be improved if the integrals are extended to arbitrary values of $m$. Analytical integrals for the first term in eq.~\ref{eq:f+} for the functional form of the photoelectric cross-section above have been provided in Appendix section~\ref{sec:Analytical}. Using the analytical integral from section \ref{sec:Analytical}, the decrement in the real part of the refractive index, $\delta$ can be parametrized using the infinite series of the form 
	\begin{center}
		\boxed{
			\delta = \frac{A_{1}}{(\hbar\omega)^{2}} + \frac{A_{2}}{(\hbar\omega)^{m}} + \frac{A_{3}}{(\hbar\omega)^{m+1}} + \frac{A_{4}}{(\hbar\omega)^{4}} + \cdots,
			\label{eq:series}
		} 
	\end{center}
	where, $A_{1}, A_{2}, A_{3} \cdots$ etc. are material dependent. 
	\begin{figure}[!ht]
		\centering
		\includegraphics[width=0.7\textwidth]{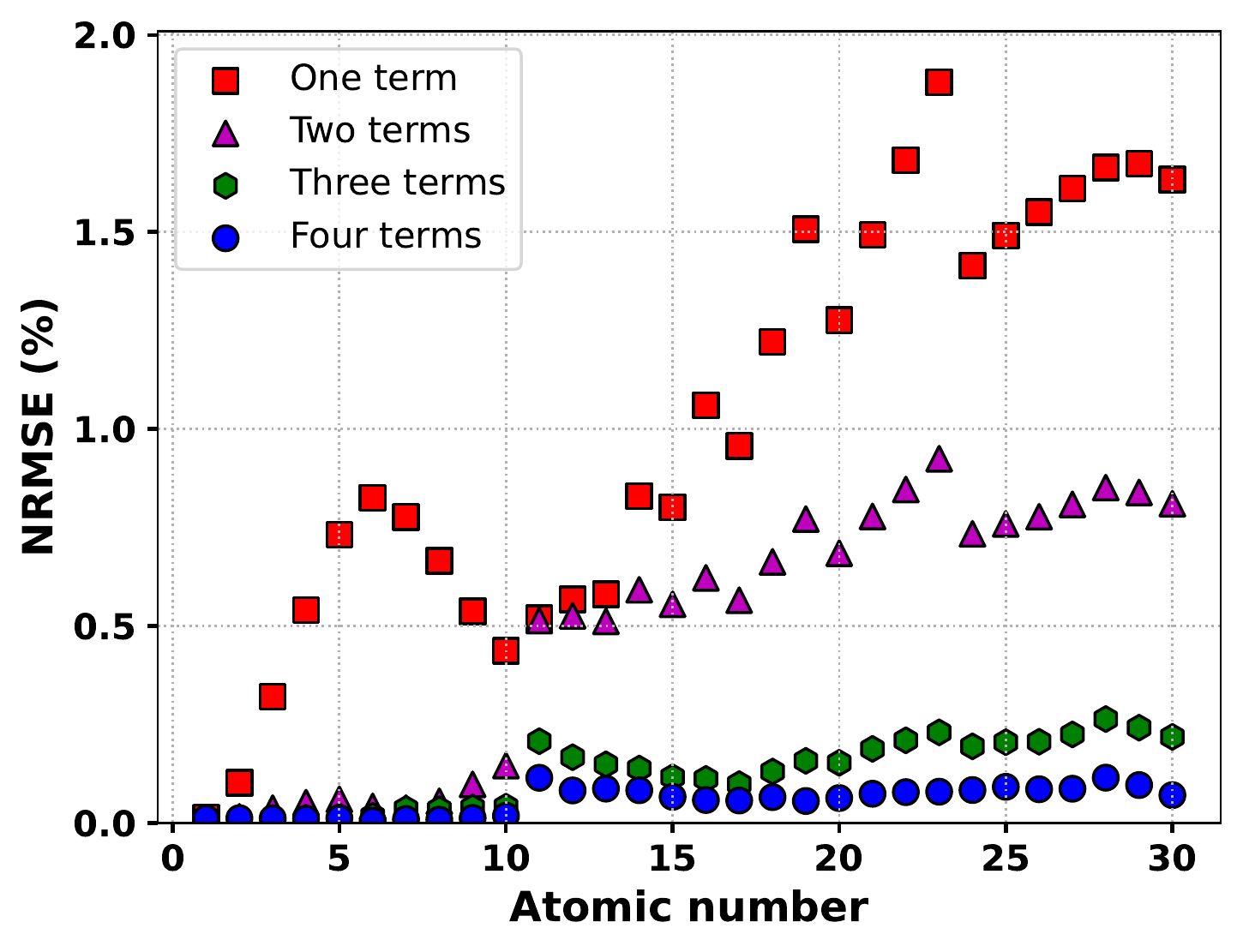}
		\caption{The mean error for best fit of analytic $\delta$ values as a function of the atomic number using eqn.~\ref{eq:series}. The energy range for the fit starts just above the K-edge energy of the elements and goes up to $\sim 400$ keV. The mean error decreases as the number of terms increases, with the mean error never exceeding $1\%$ when at least two terms are used.
			\label{fig:err_delta_Z}}
	\end{figure}
	For incident x-ray energies much higher than the K-edge of constituent elements in a sample, the approximation can be tuned to an arbitrary level of precision by including more terms in the parametrization. However, in practice, $1-2$ terms give an error $\ll 1 \%$, sufficient for most applications. Note that the other absorption edges (L1, L2, L3, M1 edges etc.) are lower energy than the K-edge. Therefore, the condition of the x-ray energies higher than the K-edge is sufficient for the series representation to be valid. Fig.~\ref{fig:err_delta_Z} shows the normalized mean squared error (NMSE) between the analytical values of $\delta$ and the best fit in the energy range starting just above the K-edge energy of the elements to $\sim 400$ keV. The fit was performed using one to four terms for atomic number up to $30$ (Zn). The NRMSE is given by the equation.
	\begin{equation}
		NRMSE (\%) = 100\times\frac{\vert\vert\mathbf{x}_{\textrm{meas}}-\mathbf{x}_{\textrm{fit}}\vert\vert_{2}}{\vert\vert\mathbf{x}_{\textrm{meas}}\vert\vert_{2}}.
	\end{equation}
	Here, $\mathbf{x}_{\textrm{meas}}$ are the analytical values, $\mathbf{x}_{\textrm{fit}}$ are the best fit values and $\vert\vert\cdot\vert\vert_{2}$ represents the $L^2$ norm.
	The accuracy of the fit improves if more terms are included, with the error never exceeding $1\%$ if at least two terms are included. When fewer terms are included, the discrepancy between the analytical and best-fit values increases for higher Z elements. However, including three or more terms keeps the error below $0.1\%$. For most quantitative PCT applications, using one or two terms of the series can provide the necessary accuracy.
	
	\section{Results\label{sec:results}}
\begin{figure}[!ht]
		\centering
		\includegraphics[width=\textwidth]{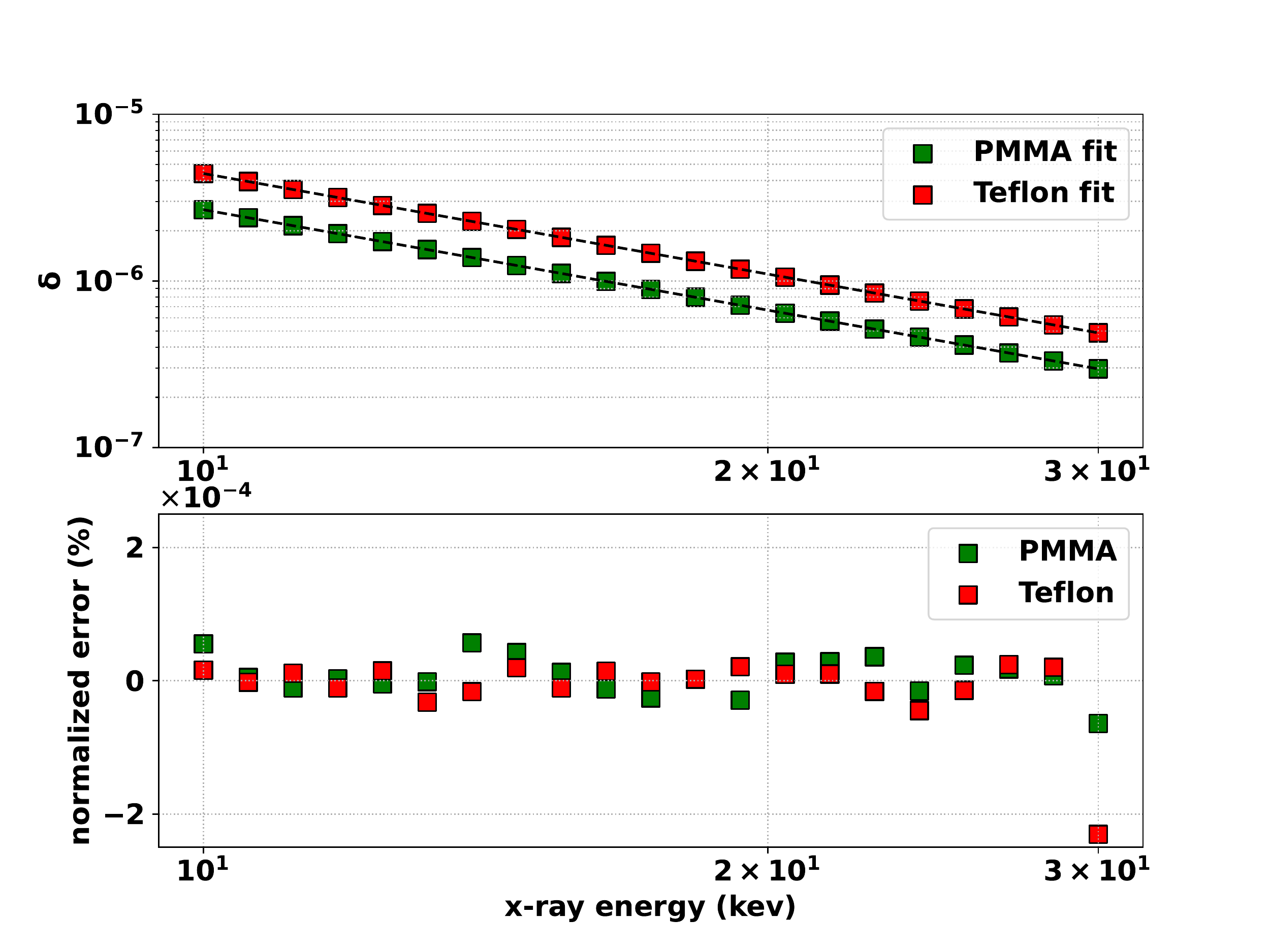}
		\caption{\textbf{Top}$\vert$ Tabulated $\delta$ values from Ref.~\cite{Henke1993} for PMMA (green squares) and Teflon (red squares). The best fit lines using the first five terms of the series is shown using the dashed line. \textbf{bottom}$\vert$ Percent error between best fit and tabulated values.}
		\label{fig:polymers}
	\end{figure}

	This section presents the results of interpolating the model and experimental data using the series expansion presented in the previous section. For the model data, we choose the refractive index decrement, $\delta$, as a function of energy up to $30$ keV from Ref. \cite{Henke1993}. Fig.~\ref{fig:polymers}, the top panel presents a log-log plot of the tabulated $\delta$ values for the polymer systems, PMMA (C$_5$O$_2$H$_8$) and Teflon (C$_2$F$_4$) using the green and red square markers respectively.
	The average atomic number was used to determine the value of the $m$ in the series representation. The black dashed lines show the best fit using the first $5$ terms of the infinite series. The normalized error (NE) for the fit, given by the following equation, is shown in the bottom panel.
	\begin{equation}
		NE (\%) = 100\times\frac{\delta_{\textrm{meas}} - \delta_{\textrm{fit}}}{\delta_{\textrm{meas}}}.
		\label{eq:error}
	\end{equation}
	Here, $\delta_{\textrm{meas}}$ is the experimental value and $\delta_{\textrm{fit}}$ is the best fit value. The mean error for the energy range is $2.2\times10^{-5} \%$ for both polymers. If only the first two terms are used, the mean error increases to approximately $0.01 \%$ over the energy range. We note that a polynomial in $\log-\log$ space also gives a gives an excellent fit to $\delta$, i.e. $\log\delta = c_0 + c_1\log E + c_2(\log E)^2+\cdots$. However, the series representation fit is consistently better for the same number of terms used than the $\log$ series representation. In addition, the series representation fit makes it easier to solve the 3D reconstruction problem since each coefficient can be 3D reconstructed independently due to linearity.
	\begin{figure}[!ht]
		\centering
		\includegraphics[width=\textwidth]{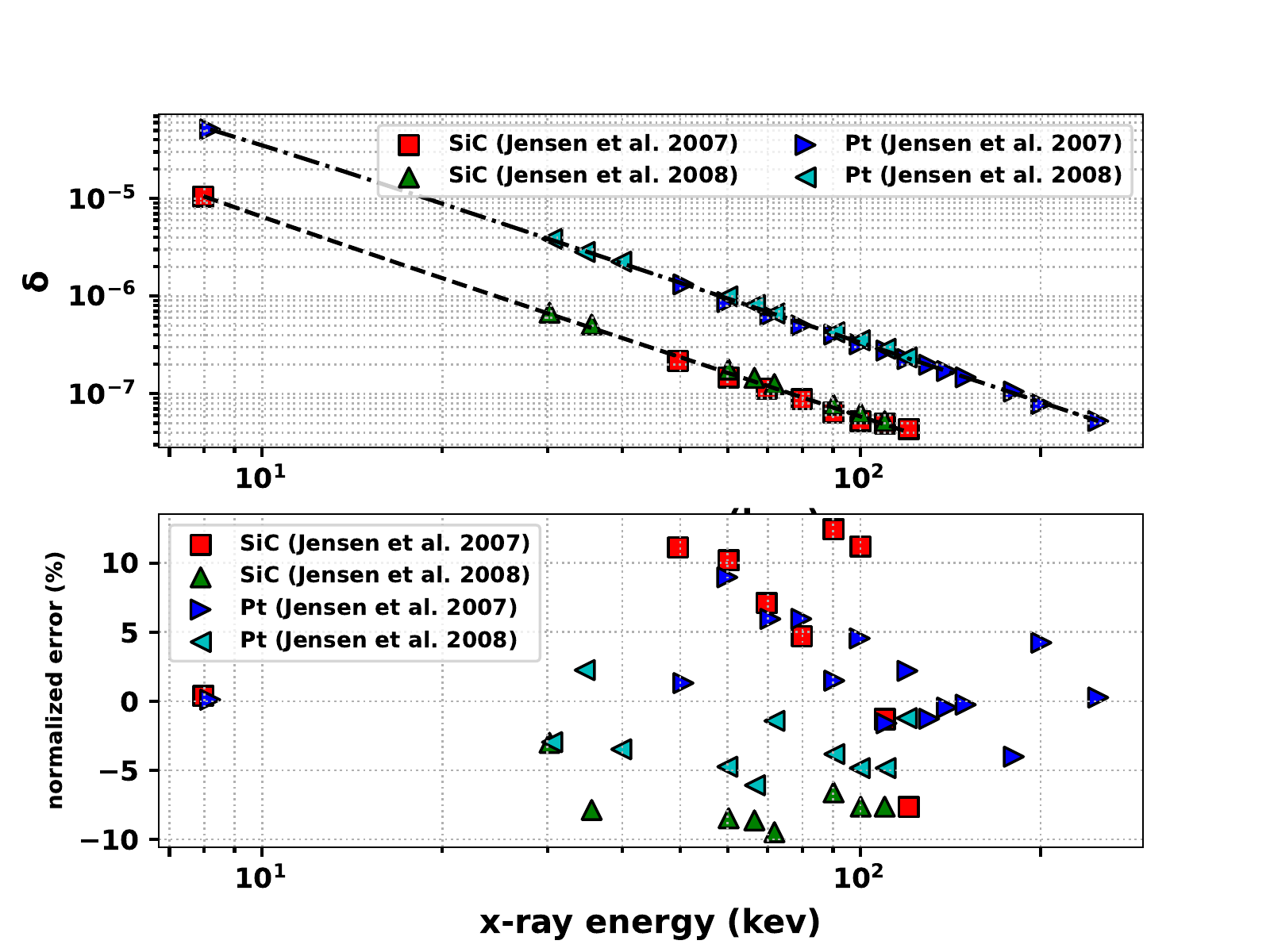}
		\caption{\textbf{Top}$\vert$ Experimentally determined $\delta$ values from Refs.~\cite{Jensen2007,Jensen2008} for SiC (red, green markers) and Pt (blue, cyan markers). The best fit lines using only two terms of the series ($1/E^{2}$ and $1/E^{m}$) is shown using the dashed line. For SiC, the value of $m$ was determined by taking the average values of $m$ for Si and C atoms. \textbf{bottom}$\vert$ Percent error between best fit and experimental values.}
		\label{fig:expt}
	\end{figure}
	Finally, we test our series representation for interpolating experimentally determined refractive index of high energy x-rays. The normalized error has the same definition as eqn.~\ref{eq:error}. The top panel in Fig.~\ref{fig:expt} shows the log-log plot experimentally determined values of the refractive index decrement as a function of x-ray energy for two material systems: SiC and Pt. Data from two separate studies \cite{Jensen2007,Jensen2008} were used for each of the materials. Only the first two terms of the series, $1/E^2$ and $1/E^{m}$ were used to interpolate the experimental values. The best-fit curves are shown by the dashed and dot-dashed lines in the plot. The bottom panel shows the percentage error between the best-fit line and the experimental values. As expected, while the series representation interpolates the experimental data well, the agreement is less accurate than tabulated values. The experimental values of the two studies show a systematic bias. The values from Ref.~\cite{Jensen2007} show positive errors from the best fit, while those from Ref.~\cite{Jensen2008} show negative errors. If interpolation is performed on only one data set, the maximum error reduces to $\sim 4-5 \%$.
	
	\section{Conclusion\label{sec:conclusions}}
	A series representation for the energy dependence of refractive index decrement, $\delta$ is derived in the high energy regime far from any absorption edges. The functional forms are physically motivated and valid over a wide energy range. The proposed functional form fits the tabulated values of $\delta$ with errors much less than $1\%$ over the energy range of $10 - 200$ keV. The same series representation can also be used to model the refractive index as a function of energy for more complex systems, such as polymers. This is demonstrated using two plastics, PMMA and Teflon. The presented series representation serves to make the problem of quantitative tomographic reconstruction of refractive index using poly-chromatic sources well-posed. 
	
	\begin{backmatter}
		
		\bmsection{Acknowledgments}
		\footnotesize{\normalfont{The research was supported by the Laboratory Directed Research and Development Program at LLNL (project no. 22-ERD-011). This work performed under the auspices of the U.S. Department of Energy by Lawrence Livermore National Laboratory under Contract No. DE-AC52-07NA27344 (LLNL-JRNL-XXXXXX).}}
		
		\bmsection{Disclosures}
		The authors declare no conflicts of interest.
		
		\bmsection{Data Availability Statement}
		
	\end{backmatter}
	
	
	\bibliography{references}
	
	\onecolumn
\noindent
{\Large{Appendix}}
\renewcommand{\thefigure}{A\arabic{figure}}

\renewcommand{\thefigure}{A\arabic{figure}}
\renewcommand{\thetable}{A\arabic{table}}
\renewcommand{\theequation}{A\arabic{equation}}
\renewcommand\thesection{A\arabic{section}}
\renewcommand\thesubsection{A\thesection.\arabic{subsection}}

\captionsetup[table]{name=Table ,labelsep=period}
\captionsetup[Table]{labelfont=bf}

\setcounter{table}{0}
\setcounter{figure}{0}
\setcounter{table}{0}
\setcounter{section}{0}
\setcounter{equation}{0}
\normalsize 

\section{Photoelectric cross-section\label{sec:app1}}
\begin{figure}[htbp]
	\centering
	\includegraphics[width=\textwidth]{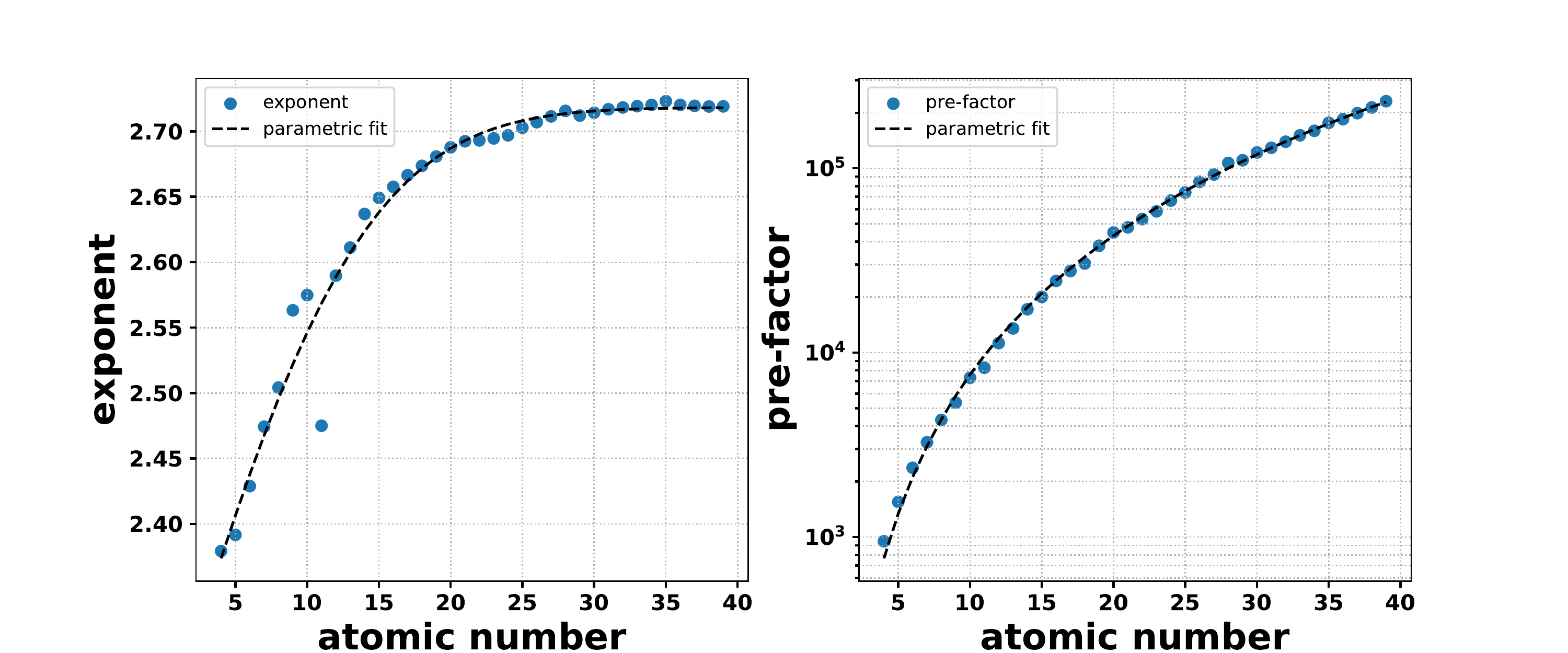}
	\caption{Functional dependence of pre-factor and exponent on the atomic number.}
	\label{fig:supp1}
\end{figure}
The photoelectric cross sections above the K-edge were fit to simple functional forms of the form $\sigma(\hbar\omega) = K/(\hbar\omega)^{m}$for all elements up to $Z = 40$. The pre-factor, $K$ and the exponent, $m$ for the different atomic numbers are shown in the Fig.~\ref{fig:supp1}. These parameters also follow simple functional forms as a function of the atomic number as
\begin{eqnarray}
	m(Z) &=& \alpha_{0} + \alpha_{1}*\textrm{erf}\left(\alpha_{2} Z\right), \\ \nonumber
	K(Z) &=& \beta_{0}Z^{n}.
	\label{eq:variation_Z_m}
\end{eqnarray}
The functional form below the K-edge and above the L1-edge remains the same, but the value of the exponent is different as compared to the above K-edge case
\begin{equation}
	\sigma(\hbar\omega) = \begin{cases}
		\frac{K^{\prime}}{\left(\hbar\omega\right)^{n}} &\quad \hbar\omega \leq \epsilon_{K}\\
		\frac{K}{\left(\hbar\omega\right)^{m}} &\quad \hbar\omega > \epsilon_{K}.
	\end{cases}
	\label{eq:pe_cross}
\end{equation}

These functional forms can be used to compute the optimal exponent in the basis functions for a given range of atomic numbers in the sample. 

\section{Analytical Integrals\label{sec:Analytical}}
James in Ref.~\cite{james1965} eqns.~4.35-4.37 has given the analytical form of the anomalous scattering form factors due to the K-edge for specific values of $n$ in equation \ref{eq:pe_cross}. In particular, the exact integrals are given for the values of $n = 2, 2.5$ and $3$. The results are tabulated here for the readers' convenience. 
\begin{equation}
	f^{\prime}_{\text{\tiny K}}\left(\hbar\omega\right) = \begin{cases} \epsilon_{\text{\tiny K}} g_{\text{\tiny K}}(\epsilon_{\text{\tiny K}})\left(\frac{1}{2\hbar\omega}\right)\log\left[\frac{\vert(1 - (\epsilon_{\text{\tiny K}}/\hbar\omega))\vert}{(1 + (\epsilon_{\text{\tiny K}}/\hbar\omega)}\right] &\quad n = 2 \\ \\
		(3g_{\text{\tiny K}}(\epsilon_{\text{\tiny K}})/2)\left(\frac{\epsilon_{\text{\tiny K}}}{\hbar\omega}\right)^{3/2}\left\{\frac{\pi}{2} - \cot^{-1}(\hbar\omega/\epsilon_{\text{\tiny K}})^{1/2} - \frac{1}{2}\log\left[\frac{1 + \sqrt{\epsilon_{\text{\tiny K}}/\hbar\omega}}{\vert 1 - \sqrt{\epsilon_{\text{\tiny K}}/\hbar\omega}\vert}\right]\right\} &\quad n = 2.5, \\ \\
		g_{\text{\tiny K}}(\epsilon_{\text{\tiny K}})\left(\frac{\epsilon_{\text{\tiny K}}}{ \hbar\omega}\right)^{2}\log\left[\frac{\vert 1 - (\epsilon_{\text{\tiny K}}/\hbar\omega)^{2} \vert}{(\epsilon_{\text{\tiny K}}/\hbar\omega)^{2}}\right] &\quad n = 3.
	\end{cases}
	\label{eq:analytic_fpp}
\end{equation}
Here, $\epsilon_{\text{\tiny K}}$ is the K-edge energy of an element and $g_{\text{\tiny K}}(\epsilon_{\text{\tiny K}})$ denotes the oscillator strength associated with the K-edge energy, $\epsilon_{\text{\tiny K}}$ and is given by the expression (Eq. 4.33 in Ref.~\cite{james1965}),
\begin{equation*}
	g_{\text{\tiny K}}(\epsilon_{\text{\tiny K}}) = \frac{1}{2\pi^2\alpha}\frac{\epsilon_{\text{\tiny K}}}{m - 1}\sigma\left({\epsilon_{\text{\tiny K}}}\right).
	\label{eq:osc_strength}
\end{equation*}
$\sigma\left({\epsilon_{\text{\tiny K}}}\right)$ is the photoelectric cross-section at the K-edge of the atom and $m$ is the exponent appearing in eqn.~\ref{eq:pe_cross}. The other symbols are fundamental constants and have been defined previously.
Since this exponent varies between $2$ and $3$ with atomic number as shown in Fig.~\ref{fig:supp1}(a), this section aims to generalize the results of James by computing the analytical integrals for any value of $n$. The asymptotic behavior of these analytical integrals gives us the basis functions to use for $\delta$ far from the absorption edge of the elements in the sample.

The integral which is the first term in equation~\ref{eq:f+} has the photoelectric cross-section. The above integral can be broken up at $x=1/2$, which corresponds to energy at the absorption edge, i.e. $\hbar\omega^{\prime}$ at $\epsilon$. 
\begin{eqnarray*}
	I &=& \frac{1}{2\pi^2\alpha}\int_{0}^{1}\frac{\sigma(\epsilon/x)\epsilon^2 - \sigma(\hbar\omega)(\hbar\omega)^2 x^2}{(\hbar\omega)^2x^{4} - \epsilon^2 x^2}\textrm{d}x, \\ \nonumber
	&=&  \frac{1}{2\pi^2\alpha}\left[\int_{0}^{1/2}\frac{\sigma(\epsilon/x)\epsilon^2 - \sigma(\hbar\omega)(\hbar\omega)^2 x^2}{(\hbar\omega)^2x^{4} - \epsilon^2 x^2}\textrm{d}x + \int_{1/2}^{1}\frac{\sigma(\epsilon/x)\epsilon^2 - \sigma(\hbar\omega)(\hbar\omega)^2 x^2}{(\hbar\omega)^2x^{4} - \epsilon^2 x^2}\textrm{d}x\right].
\end{eqnarray*}
This is done to account for the different behavior of the photoelectric cross section above and below the edge. The first term in the above equation is the above K-edge case the the second term is below that value. Above the K-edge, the photoelectric cross-section has the form $\sigma(E) = K/E^{m}$, while below the K-edge, the cross-section has the form $\sigma(E) = K^{\prime}/E^{n}$. Below the K-edge, the integral can be broken down into smaller intervals between K-edge, L1, L2, L3, M1 edges etc. However, that does not alter the overall behavior discussed in the rest of this section. Substituting the functional forms in the integral above,
\begin{eqnarray*}
	I &=& \frac{1}{2\pi^2\alpha}\left[\frac{K}{(\hbar\omega)^{2}\epsilon^{m-2}}\int_{0}^{1/2}\frac{x^{m} - (\epsilon/\hbar\omega)^{(m-2)}x^{2}}{x^{4} - (\epsilon/\hbar\omega)^2 x^2}\textrm{d}x \right. \\ \nonumber
	&+& \left. \frac{K^{\prime}}{(\hbar\omega)^{2}\epsilon^{n-2}} \int_{1/2}^{1}\frac{x^{n} - \frac{K\epsilon^{n-2}}{K^{\prime}(\hbar\omega)^{m-2}} x^2}{x^{4} - (\epsilon/\hbar\omega)^2 x^2}\textrm{d}x\right].
	\label{eq:integral}
\end{eqnarray*}
The integrals above were calculated using \textit{Mathematica 13} \cite{Mathematica}. The integral results are composed of the special hypergeometric function, ${}_2F_{1}\left(a, b; c; z\right)$ \cite{hypergeometric}. Since we are only interested in the above K-edge cases, i.e. $\hbar\omega \gg \epsilon$, the functional behavior of the integral can be obtained by the series representation of the functions in the limit $\epsilon/\hbar\omega \ll 1$. The first term, $I_{1}$ in eq.~\ref{eq:integral} is given by
\begin{eqnarray*}
	I_{1} &=& \frac{K}{(\hbar\omega)^{2}\epsilon^{m-2}}\int_{0}^{1/2}\frac{x^{m} - (\epsilon/\hbar\omega)^{(m-2)}x^{2}}{x^{4} - (\epsilon/\hbar\omega)^2 x^2}\textrm{d}x \\ \nonumber  
	&=& \frac{K}{(\hbar\omega)^{2}\epsilon^{m-2}} \left\{ \frac{(\epsilon/\hbar\omega)^{m-3}}{2}\left\{ \log\left(\frac{1-(2\epsilon/\hbar\omega)}{1+(2\epsilon/\hbar\omega)}\right) + \pi\tan \left(\frac{m\pi}{2}\right)\right\} \right. \\ \nonumber
	&+& \left.\frac{2^{-m+3}}{m-3}{}_2F_{1}\left(1, \frac{-m+3}{2}; \frac{-m+5}{2}; (2\epsilon/\hbar\omega)^{2}\right) \right\}.
\end{eqnarray*}
In the limit $\epsilon/\hbar\omega \ll 1$, the series expansion of this term is given by the following polynomial in $(\epsilon/\hbar\omega)$,
\begin{eqnarray*}
	I_{1} &=& \frac{K}{\epsilon^{m}}\left\{ \frac{\pi}{2}\tan\left(\frac{m\pi}{2}\right) \left(\frac{\epsilon}{\hbar\omega}\right)^{m-1} + \frac{2^{-m+3}}{m-3}\left(\frac{\epsilon}{\hbar\omega}\right)^{2} + 2\left(\frac{\epsilon}{\hbar\omega}\right)^{m} + \frac{2^{-m+5}}{m - 5}\left(\frac{\epsilon}{\hbar\omega}\right)^{4} + \frac{8}{3}\left(\frac{\epsilon}{\hbar\omega}\right)^{m+2} + \cdots \right\}
\end{eqnarray*}
Since the exponent, $m$ for light elements is in the range $2.4 - 2.7$ as shown in Fig.~\ref{fig:supp1}(a), the above equation is in arranged in increasing power of $(\epsilon/\hbar\omega)$. Similarly, the second term in the eq.~\ref{eq:integral} is given by the following equation
\begin{eqnarray*}
	I_{2} &=& \frac{K^{\prime}}{(\hbar\omega)^{2}\epsilon^{n-2}} \int_{1/2}^{1}\frac{x^{n} - \frac{K\epsilon^{n-2}}{K^{\prime}(\hbar\omega)^{m-2}} x^2}{x^{4} - (\epsilon/\hbar\omega)^2 x^2}\textrm{d}x \\ \nonumber
	&=& \frac{-K^{\prime}}{(\hbar\omega)^{2}\epsilon^{n-2}(n-3)}\left\{ {}_2F_{1}\left(1, \frac{-n+3}{2}; \frac{-n+3}{2}; (\epsilon/\hbar\omega)^{2}\right) -2^{-n+3} {}_2F_{1}\left(1, \frac{-n+3}{2}; \frac{-n+3}{2}; (2\epsilon/\hbar\omega)^{2}\right) \right. \\ \nonumber
	&-& \left. \frac{\kappa}{2}\left(\frac{\epsilon}{\hbar\omega}\right)^{m-3}\log\left(\frac{(-1 + (\epsilon/\hbar\omega))(1 + (2\epsilon/\hbar\omega))}{(-1 + (2\epsilon/\hbar\omega))(1 + (\epsilon/\hbar\omega))}\right)\right\}
\end{eqnarray*}
Here, $\kappa = K\epsilon^{n-m}/K^{\prime}$. As with the first term, $I_{1}$, in the limit $\epsilon/\hbar\omega \ll 1$, the series expansion of this term is given by the following polynomial in $(\epsilon/\hbar\omega)$,
\begin{eqnarray*}
	I_{2} &=& \frac{K^{\prime}}{\epsilon^{n}}\left\{ -\kappa \left(\frac{\epsilon}{\hbar\omega}\right)^{m-2} + \frac{1 - 2^{-n+3}}{n - 3}\left(\frac{\epsilon}{\hbar\omega}\right)^{2} + -\frac{7\kappa}{3}\left(\frac{\epsilon}{\hbar\omega}\right)^{m} + \frac{1 - 2^{-n+5}}{n - 5}\left(\frac{\epsilon}{\hbar\omega}\right)^{4} + \cdots -\frac{31\kappa}{5}\left(\frac{\epsilon}{\hbar\omega}\right)^{m+2}\right\}
\end{eqnarray*}
Summing up $I_{1}$ and $I_{2}$ results in the following functional form of the overall integral, $I$  in increasing power of the negative exponent of $\hbar\omega$
\begin{eqnarray*}
	I &=& \frac{-\kappa K^{\prime}}{\epsilon^{n+2-m}} \left(\frac{1}{\hbar\omega}\right)^{m-2} + \frac{\pi K}{2\epsilon}\tan\left(\frac{m\pi}{2}\right)\left(\frac{1}{\hbar\omega}\right)^{m-1} + \left(\frac{2^{-m+3}K}{(m-3)\epsilon^{m-2}} + \frac{(1 - 2^{-n+3})K^{\prime}}{(n - 3)\epsilon^{n-2}}\right) \left(\frac{1}{\hbar\omega}\right)^{2} \\ \nonumber
	&+& \left(2K - \frac{7\kappa}{3\epsilon^{n-m}}\right)\left(\frac{1}{\hbar\omega}\right)^{m} + \cdots.
\end{eqnarray*}
The relationship between the anomalous scattering factor in the series above, and the decrement in the real part of the refractive index, $\delta$ is given by eqn.~\ref{eq:deltabeta}. The relationship introduces an extra factor of $(1/\hbar\omega)^{2}$. The series expansion for $\delta$ is then given by
\begin{eqnarray*}
	\delta &=& \frac{2\pi \rho_{n} r_e}{(\hbar\omega)^2} \left(Z + \frac{-\kappa K^{\prime}}{\epsilon^{n+2-m}} \left(\frac{1}{\hbar\omega}\right)^{m-2} + \frac{\pi K}{2\epsilon}\tan\left(\frac{m\pi}{2}\right)\left(\frac{1}{\hbar\omega}\right)^{m-1} + \left(\frac{2^{-m+3}K}{(m-3)\epsilon^{m-2}} + \frac{(1 - 2^{-n+3})K^{\prime}}{(n - 3)\epsilon^{n-2}}\right) \left(\frac{1}{\hbar\omega}\right)^{2} \right. \\ \nonumber
	&+& \left. \left(2K - \frac{7\kappa}{3\epsilon^{n-m}}\right)\left(\frac{1}{\hbar\omega}\right)^{m} + \cdots \right).
\end{eqnarray*}
Inspecting the above equation, the functional dependence of $\delta$ is captured by the functions $(1/\hbar\omega)^2, (1/\hbar\omega)^m, (1/\hbar\omega)^{m+1}, (1/\hbar\omega)^4, \cdots$.
	
\end{document}